\documentclass[preprint]{aastex}
\usepackage{graphicx}
\usepackage{subfigure}
\usepackage{mathrsfs,amssymb}
\usepackage{amsmath}
\usepackage{natbib}
\usepackage{color}
\usepackage{ulem}
\usepackage{bm}
\usepackage{multirow}

\def\gtsima{$\; \buildrel > \over \sim \;$}
\def\ltsima{$\; \buildrel < \over \sim \;$}
\def\gsim{\lower.5ex\hbox{\gtsima}}
\def\lsim{\lower.5ex\hbox{\ltsima}}
\def\simgt{\lower.5ex\hbox{\gtsima}}
\def\simlt{\lower.5ex\hbox{\ltsima}}
\def\simpr{\lower.5ex\hbox{\prosima}}
						
\begin{document}

\title{{\Large The brief era of direct collapse black hole formation}}

\author{
Bin Yue\altaffilmark{1},
Andrea Ferrara\altaffilmark{1,2},
Ruben Salvaterra\altaffilmark{3},
Yidong Xu\altaffilmark{4},
Xuelei Chen\altaffilmark{4,5}
}
\altaffiltext{1}{Scuola Normale Superiore, Piazza dei Cavalieri 7, I-56126 Pisa, Italy}
\altaffiltext{2}{Kavli IPMU (WPI), Todai Institutes for Advanced Study, the University of Tokyo}
\altaffiltext{3}{INAF, IASF Milano, via E. Bassini 15, I-20133 Milano, Italy}
\altaffiltext{4}{National Astronomical Observatories, Chinese Academy of Sciences,
20A Datun Road, Chaoyang, Beijing 100012, China}
\altaffiltext{5}{Center of High Energy Physics, Peking University, Beijing 100871, China}
		
\begin{abstract}
It has been proposed that the first, intermediate-mass ($\approx 10^{5-6}~M_\odot$) black holes might form through direct collapse
of unpolluted gas in atomic-cooling halos exposed to a strong Lyman-Werner (LW) or near-infrared (NIR) radiation. As these systems are expected to be Compton-thick, photons above 13.6 eV are largely absorbed and re-processed into lower energy bands. It follows that direct collapse black holes (DCBHs) are very bright in the LW/NIR bands, typically outshining small high-redshift galaxies by more than 10 times. Once the first DCBHs form, they then trigger a runaway process of further DCBH formation, producing a sudden rise in their cosmic mass density. The universe enters the ``DCBH era" at $z \approx 20$ when a large fraction of  
atomic-cooling halos are experiencing DCBH formation. By combining the clustering properties of 
the radiation sources with Monte Carlo simulations we show that in this scenario the DCBH mass 
density rises from $\sim 5$~$M_\odot$ Mpc$^{-3}$ at $z\sim 30$ to the peak value $\sim5\times10^5 M_\odot$ Mpc$^{-3}$ at $z \sim 14$ in our fiducial model. However, the abundance of \textit{active} (accreting) DCBHs drops after $z \sim 14$, as gas in the potential formation sites (unpolluted halos with virial temperature slightly above $10^4$~K) is photoevaporated.
This effect almost completely suppresses DCBH formation after $z\sim 13$. 
The DCBH formation era lasts only $\approx 150$ Myr, but it might crucially provide the seeds of the
supermassive black holes (SMBHs) powering $z\sim6$ quasars.
\end{abstract}


\maketitle

\section{INTRODUCTION}

The formation process of the intermediate-mass black holes (BHs) through direct collapse has been investigated 
by numerical and theoretical studies
\citep{2002ApJ...569..558O,2003ApJ...596...34B,
2004MNRAS.354..292K,	
2006MNRAS.371.1813L,2006MNRAS.370..289B,
2008MNRAS.387.1649B,
2009MNRAS.396..343R,2010MNRAS.402..673B,
2012ApJ...750...66J,2012arXiv1211.0548J,2013MNRAS.433.1607L,2013MNRAS.436.2989L}.
Such BHs, due to their much larger mass than the BHs form 
after the death of Pop III stars (typically several tens $M_\odot$, see \citealt{2011Sci...334.1250H}),
are good candidates of the seeds of the supermassive black holes (SMBHs) which 
were observed at $z\sim6$ \citep{2001AJ....122.2833F}. In this direct collapse
black hole (DCBH) scenario, the formation of a DCBH requires
a host halo: (a) with virial 
temperature above $10^4$~K; (b) exposed to strong Lyman-Werner (LW, photons with 
energy between 11.2 eV and 13.6 eV) radiation or near-infrared (NIR) radiation 
with energy above $0.76$~eV \citep{2007ApJ...665L..85C,2010MNRAS.402.1249S}.
Under these conditions, the $\rm H_2$ can no longer form and the gas 
in this halo can only cool through Ly$\alpha$ emission, 
leading to a quasi-isothermal contraction without fragmentation
and finally allowing  a BH seed with mass $\sim10^4 - 10^6~M_\odot $ to form at the center of the halo.

If such DCBHs largely formed at $z\simgt 13$ and are all 
Compton-thick, with a typical mass $\sim10^6~M_\odot$ and typical accretion time scale
$\sim 30~$Myr, they might be excellent candidate sources of the fluctuations of the source-subtracted 
near infrared background (NIRB)
\citep{2005Natur.438...45K,2007ApJ...654L...5K, 
2007ApJ...657..669T,2011ApJ...742..124M,2012ApJ...753...63K,2012Natur.490..514C}
which is hard attributed to normal galaxies \citep{2012ApJ...752..113H,2012ApJ...756...92C,
2013MNRAS.431..383Y}. 
As proposed in \citet{2013MNRAS.433.1556Y},
DCBHs
also naturally generate a cosmic X-ray background (CXB) that is correlated
with the NIRB, which is observed by \citet{2013ApJ...769...68C}.

The minimum LW intensity necessary to sustain the formation of the DCBH has been somewhat
debated in recent literatures. Early studies favored values as high as $\sim10^3$ (in units of 
$10^{-21}$erg s$^{-1}$cm$^{-2}$Hz$^{-1}$sr$^{-1}$) for a $10^4$~K blackbody spectrum,
which represents a typical spectrum of Pop II stars;
and $\sim10^{4-5}$ for a $10^5$~K blackbody spectrum,
which characterizes metal-free stars
\citep{2000MNRAS.317..175S,2001ApJ...546..635O,2002ApJ...569..558O,2003ApJ...596...34B}. 
The critical intensity for the $10^4$~K blackbody spectrum is lower because in this case,
the main mechanism that quenches the $\rm H_2$ formation is not to directly dissociate the
$\rm H_2$ themselves, but to dissociate the 
$\rm H^{-}$ \citep{2007ApJ...665L..85C,2012MNRAS.425L..51W} 
\--- the necessary intermediary for the $\rm H_2$ formation \--- through the NIR radiation 
above $0.76$~eV.
However, a more detailed work by \citet{2010MNRAS.402.1249S} revised the value 
downward: e.g. for a $10^4$~K blackbody spectrum, it is $\sim 30-300$. 
The critical LW intensity could be even lower if a more refined treatment of self-shielding factor is implemented (\citealt{2011MNRAS.418..838W}, Z. Haiman, personal communication). 

Usually the probability for a halo to be embedded in a radiation field with 
super-critical LW intensity\footnote{Throughout this paper, when we are talking about the 
``super-critical LW intensity", we refer to either the LW radiation above a critical 
intensity \--- in case the $\rm H_2$ cooling is suppressed due to the direct dissociation of $\rm H_2$, or 
super-critical NIR intensity which can be characterized by the 
intensity at LW band when the spectrum is already known \--- in case the $\rm H^-$ dissociation
plays the main role.
}
is very small if normal galaxies and/or Pop III stars are the only sources of LW photons \citep{2008MNRAS.391.1961D,2012MNRAS.425.2854A,2013MNRAS.432.3438A}; in this case  DCBHs are very rare. 
However, \citet{2013MNRAS.433.1556Y} pointed out that a Compton-thick BH could be very luminous below 13.6 eV due to the  reprocessed nebular emission. This means that, once the first DCBHs form, they could ignite a runaway process  favoring the formation of more DCBHs and leading to a steep rise in the DCBH abundance; we dub this well-defined period during cosmic evolution as the ``DCBH era'', lasting for about 150 Myr.

The probability that a halo sees a LW intensity above the critical value can be  calculated by Monte Carlo simulations \citep{2008MNRAS.391.1961D}, in which the LW radiation from previously formed DCBHs is explicitly taken into account. With this probability, and the probability of a halo being unpolluted, we get the formation rate of the DCBH and build the picture of the rise process of its abundance.    
The process comes to an end when the host atomic-cooling halos ($T_{\rm vir} \simgt 10^4$~K) are rapidly
photoevaporated by ionizing photons. Photoevaporation drastically reduces the DCBH formation rate 
and almost completely quenches their formation for  $z\simlt 13$. 
This is the basic idea of this work.
We aim at building a solid picture of the rapid rise (and fall) of the DCBH formation rate during the DCBH era.

The layout of this paper\footnote{Throughout the paper, we use $M_{\rm T3}$, $M_{\rm T4}$ and $M_{\rm 2T4}$ to denote the mass of halos with $T_{\rm vir} = 10^3$~K, $10^4$~K and $2\times10^4$~K respectively at the corresponding redshift. The cosmological parameters are the same as used in \cite{2011MNRAS.414..847S}:  $\Omega_m$=0.26, $\Omega_\Lambda$=0.74, $h$=0.73, $\Omega_b$=0.041, $n=1$ and $\sigma_8$=0.8. We use the the transfer function in \cite{1998ApJ...496..605E}.}
 is as follows: in Sec. \ref{methods} we introduce the clustering of the spatial distribution 
of halos at high redshifts, the LW luminosity of these sources and the various mechanisms that influence the 
formation of DCBHs. We give results in Sec. \ref{results}, the conclusions are presented in Sec. \ref{conclusions}. 
A list of the mostly used symbols is given in Tab. \ref{symbols}.
The calculation of the probability that 
a halo sees a super-critical LW intensity is introduced in Appendix \ref{AppA}.

\begin{table}
\begin{center}
\caption{
{List of the most used symbols.}
}
\begin{tabular}{ll} 
\hline \hline
Symbol & Definition \\
\hline
{\small $ L_{\rm LW}^{\rm gal}(M)$} &\multirow{1}{13cm}{\small LW luminosity of a normal galaxy hosted by a halo of mass $M$} \\
{\small $L_{\rm LW}^{\rm BH}(M_{\rm BH})$}&\multirow{1}{13cm}{\small LW luminosity of a DCBH with mass $M_{\rm BH}$}\\ 
{\small $l_{\rm LW}^{\rm pop3}$} &\multirow{1}{13cm}{\small LW luminosity (per stellar mass) of Pop III stars }\\ 
{\small $J_{\rm LW}^{\rm bg}$}&\multirow{1}{13cm}{\small Background LW intensity} \\
{\small $J_{\rm LW}^{\rm pop3}$}&\multirow{1}{13cm}{\small Background LW intensity from Pop III stars}\\		
{\small $J_{\rm LW}^{\rm gal}$}&\multirow{1}{13cm}{\small Background LW intensity from normal galaxies }\\
{\small $J_{\rm LW}^{\rm BH}$} & \multirow{1}{13cm}{\small Background LW intensity from DCBHs }\\
{\small $J_{\rm LW}^{\rm crit}$}&\multirow{1}{13cm}{\small Critical LW intensity for DCBH formation} \\
{\small $f_\star^{\rm gal}$} &\multirow{1}{13cm}{\small Star formation efficiency of normal galaxies}\\
{\small $f_\star^{\rm pop3}$} &\multirow{1}{13cm}{\small Star formation efficiency of Pop III stars} \\
{\small $M_{\rm crit,0}^{\rm pop3}$}&\multirow{1}{13cm}{\small Critical minihalo mass for Pop III stars 
formation without LW feedback}\\
{\small $M_{\rm crit}^{\rm pop3}$}&\multirow{1}{13cm}{\small Critical minihalo mass for Pop III stars 
formation with LW feedback}\\
{\small $M_{\rm min}^{\rm ion}$}&\multirow{1}{13cm}{\small Critical halo mass for galaxy/DCBH 
formation under photoevaporation feedback} \\
{\small $\rho_{\rm pop3}$}&\multirow{1}{11cm}{\small Mass density of Pop III stars }\\
{\small $\rho_{\rm BH}$}&\multirow{1}{11cm}{\small Mass density of active DCBHs}\\
{\small $\rho_{\rm BH}^{\rm cum}$}&\multirow{1}{11cm}{\small Cumulative DCBH mass density}\\
{\small $p_{\rm g}$}&\multirow{1}{11cm}{\small Probability of genetic enrichment} \\
{\small $p_{\rm J}$}&\multirow{1}{11cm}{\small Probability that a halo sees a super-critical LW intensity}\\
{\small $p_{\rm W}$}&\multirow{1}{13cm}{\small Probability that a halo being enriched by
metals carried by galactic winds} \\
{\small $p_{\rm J-W}$}&\multirow{2}{13cm}{\small Probability that a halo sees a super-critical 
LW intensity but without being polluted by 
galactic winds} \\
\smallskip\\
{\small $P(>z,M|M_0,z_0)$}&\multirow{2}{13cm}{\small For a halo with mass $M_0$ at $z_0$, the probability that 
its most massive progenitor with mass $M$ formed before $z$}
\smallskip \\
\smallskip \\
{\small $\frac{dP}{dz}(z,M|M_0,z_0)$}&\multirow{1}{13cm}
{\small Redshift derivative of above probability}\\
\smallskip \\
\hline
\hline
\end{tabular}
\label{symbols}
\end{center}
\end{table}

\section{METHOD}\label{methods}

\subsection{Positive feedback}\label{posfeedback}

According to the definition of the ``two-point correlation function", for a given halo of mass $m$ the number of 
surrounding halos, $N(m,z,M,l)$, with mass between [$M$, $M+dM$] and at a proper distance in [$l$, $l+dl$] is
\begin{equation}
NdMdl=4\pi l^2 (1+\xi) (1+z)^3\frac{dn}{dM}dMdl,
\label{dN}
\end{equation}
where $dn/dM$ is the halo mass function \citep{1999MNRAS.308..119S,2001MNRAS.323....1S},
$\xi(m,z,M,l)$ is the two-point correlation function 
for two halos with mass $m$ and $M$ respectively and separated by a 
proper distance $l$ at redshift $z$.
To compute $\xi(m,z,M,l)$ we use the formulae obtained in \citet{2003MNRAS.341...81I} (see also 
\citealt{2002ApJ...571..585S}); these formulae, based on the excursion set formalism, include the 
non-linear clustering of halos. We set $\xi = 0$ when $l$ is smaller than the 
sum of the virial radii of the two halos.

\begin{figure}
\begin{center}
\includegraphics[scale=0.4]{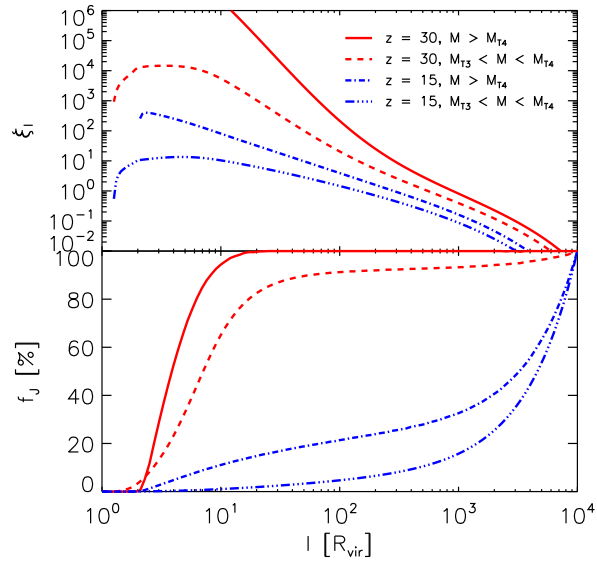}
\caption{{\it Upper}: Two-point correlation function,  $\xi_l$, averaged over surrounding halos 
in different mass ranges at $z=15,30$ for a central halo with $m = M_{\rm T4}$, as a function of 
physical distance $l$ in units of the central halo virial radius.
{\it Bottom}: Fractional contribution of surrounding sources at distance $<l$ to the radiation intensity 
as seen by the central halo (Eq. \ref{fJ}). In this figure 
we adopt $l_{\rm max} = 
10^4R_{\rm vir}$, where $R_{\rm vir}$ is the virial radius of the central halo at the corresponding redshift.}  
\label{cfl}
\end{center}
\end{figure}

The $\xi(m,z,M,l)$ averaged over all surrounding halos 
in the mass range [$M_{\rm min}$, $M_{\rm max}$] is
\begin{equation}
\xi_l(m,z,l)=\frac{1}{\bar{n}}\int_{M_{\rm min}}^{M_{\rm max}}\xi(m,z,M,l)\frac{dn}{dM}dM,
\end{equation}
where 
\begin{equation}
\bar{n}=\int_{M_{\rm min}}^{M_{\rm max}}\frac{dn}{dM}dM
\end{equation}
is the mean number density of halos between [$M_{\rm min}$, $M_{\rm max}$].
For a reference halo of mass $m = M_{\rm T4}$, we show 
$\xi_l$ as a function of $l$ at $z = 30$ and $z = 15$ in the upper panel of Fig. \ref{cfl}.
We consider separately the mass range [$M_{\rm T3}$, $M_{\rm T4}$],
representing minihalos harboring Pop III stars, 
and $>M_{\rm T4}$, corresponding to the hosts of normal galaxies or DCBHs.

At high redshifts, halos with mass above $M_{\rm T4}$ are still rare ($\sim 10^{-3}$~Mpc$^{-3}$ at $z = 30$).
However, such halos are strongly clustered, i.e. they are likely to be surrounded by close neighbors. For example, 
as shown in the upper panel of Fig. \ref{cfl}, at $z = 30$ at a distance $\approx 10\times$ the central halo virial radius, $R_{\rm vir}$,  the number density of surrounding halos with mass between [$M_{\rm T3}$, $M_{\rm T4}$] is already $\approx10^4 \times$ the mean cosmic density. For more massive surrounding halos with $M>M_{\rm T4}$ the excess is further enhanced by two orders of magnitude.
It is natural then to expect that,
as a result of the local halo density enhancement,	 
the radiation intensity as seen by the central halo overwhelmingly exceeds
the mean cosmic background intensity.

As an estimate, we can consider that the contribution of surrounding halos to the specific intensity of the radiation as seen by the central halo is 
\begin{equation}
\propto \int \frac{dl}{4\pi l^2}\int N(m,z,M,l)dM = 
\int dl (1+\xi_l)\bar{n}.
\end{equation}
The fractional contribution from all surrounding sources in the mass range [$M_{\rm min}$, $M_{\rm max}$] and 
with distance $<l$ is
\begin{equation}
f_{J}(<l, m,z)=\left.\int_{0}^{l}dl(1+\xi_l)\middle/
\int_{0}^{l_{\rm max}}dl(1+\xi_l)\right..
\label{fJ}
\end{equation}
For $m = M_{\rm T4}$, we plot $f_J$ as a function of $l$
at $z = 30$ and $z = 15$ in the bottom panel of Fig. \ref{cfl}.
The results shown in the bottom panel are for the same mass ranges as for the upper one
with the same line styles.The transition from the ``close neighbors'' dominating 
regime (note however that the real number of close neighbors fluctuates considerably
around the mean number given by $\xi$) 
to the one dominated by ``distant sources'' as redshift decreases is evident.  
Considering the radiation received by a central $m = M_{\rm T4}$ halo at $z = 30$ 
from halos above $M_{\rm T4}$ at a distance $< 10^4 R_{\rm vir}$ (roughly corresponding to the mean free path of LW photons set by Ly$\beta$ absorption), more than 90\% 
of it comes from close neighbors within $10 R_{\rm vir}$, as at such high redshift the background sources are still rare. 
Even considering only sources in 
the lower mass range [$M_{\rm T3}$, $M_{\rm T4}$], this fraction still reaches about 60\%. 
However, at $z = 15$ the fraction decreases to about 10\% and 1\% if surrounding sources are
in above two mass ranges respectively.

The highly clustered spatial distribution of halos helps to build a very inhomogeneous radiation field at high redshifts. The radiation intensity as seen by some halos is much higher than the mean cosmic background intensity. This fact creates extremely favorable (although short lasting) conditions for triggering the formation of the DCBH by the strong LW radiation from the close neighbors. 

When calculating the LW radiation as seen by atomic-cooling halos with mass $> M_{\rm T4}$ and assessing whether DCBHs could form therein, we consider two kinds of sources contributing to the LW radiation: DCBHs and normal galaxies (stars in halos with mass $> M_{\rm T4}$). Their LW luminosities are denoted by $L_{\rm LW}^{\rm BH}$ and $L_{\rm LW}^{\rm gal}$ respectively. In principle it would be necessary to include also the contribution from Pop III stars in surrounding minihalos ($< M_{\rm T4}$) when computing the impinging radiation spectrum/intensity.
However, for the following reason it is unnecessary here.
The radiation of Pop III stars is characterized by a hard spectrum\footnote{Interestingly, the slope of the spectrum of the Compton-thick BH is closer to that of normal galaxies between $\sim$3 and 10 eV,
i.e., the range probed by near-infrared observations on Earth, see \citet{2013MNRAS.433.1556Y}.}
that easily prevents $\rm H_2$ cooling by dissociating these molecules directly
and eventually suppresses the formation of Pop III stars in minihalos; 
while we include radiation from Pop III stars to calculate the LW feedback on the formation of 
Pop III stars themselves (see Sec. \ref{genetic}), its role is irrelevant in triggering the formation
of DCBHs
due to requiring a very high LW critical intensity, as
found by \citet{2012MNRAS.425.2854A} who explored a wide range of conditions.
Therefore we simply ignore the radiation from Pop III stars here. 
The spectrum of both normal galaxies and DCBHs is softer than the 
Pop III stars spectrum, the dissociation of $\rm H^-$ by NIR photons 
plays a significant role in this case, even we still use the radiation intensity
at the LW band as a criterion, it should be understood as the effects of NIR photons
characterized by the LW intensity, as the spectrum is already known.

Regarding the LW luminosity of the DCBH, we use the Spectral Energy Distribution (SED) of Compton-thick BHs derived in \citet{2013MNRAS.433.1556Y} (see the bottom panel of Fig. 1 in that paper). The luminosity depends on the BH mass, $M_{\rm BH}$, which grows following the law $M_{\rm BH}=M_{\rm BH,seed} {\rm exp}({t^\prime/t_{\rm Edd}})$, where $M_{\rm BH,seed}$ is the mass of the BH seed, $t_{\rm Edd} \approx 45$~Myr is the Eddington time scale \citep{1964ApJ...140..796S,2007ApJ...665..107P}.

At any given time, the \textit{active} DCBH population is composed by BHs formed at a time $<t_{\rm QSO}$ before; then the mass distribution follows 
\begin{equation}
\frac{dN}{dM_{\rm BH}}\propto {\rm exp}{(-t^\prime/t_{\rm E})}\frac{dn}{dt^\prime}~~~~(t^\prime \le t_{\rm QSO}),
\label{dNBH}
\end{equation}
and $dN/dM_{\rm BH} = 0$ if $t^\prime > t_{\rm QSO}$;
here $t^\prime={\rm ln}({M_{\rm BH}}/M_{\rm BH,seed})t_{\rm E}$, 
$dn/dt^\prime$ is the DCBH formation rate $t^\prime$ before
the considered time. 
Throughout this paper we use $M_{\rm BH,seed} = 10^{5.85}~M_\odot$ 
and $t_{\rm QSO} = 10^{1.48}$~Myr \citep{2013MNRAS.433.1556Y}.

For star-forming, normal galaxies, their LW luminosity is computed using 
{\tt Starburst99}\footnote{http://www.stsci.edu/science/starburst99/docs/default.htm} 
\citep{1999ApJS..123....3L,2005ApJ...621..695V,2010ApJS..189..309L}.
In the ``instantaneous burst" mode (which is more appropriate for  star formation in dwarf galaxies),
for a Salpeter stellar Initial Mass Function (IMF) \citep{1955ApJ...121..161S} in the mass range 
$0.1 - 100~M_\odot$ and metallicity $0.02~Z_\odot$, more than 60\% of the LW 
photons are emitted within 10 Myr from the burst; the mean emission rate is 
$1.3\times10^{46}$~s$^{-1}M_\odot^{-1}$, and 3600 LW photons are produced per baryon into stars during this time interval.
Therefore the LW luminosity of a halo with mass $M$ is 
\begin{align}
L_{\rm LW}^{\rm gal} &= 1.3\times10^{46} \bar{E}\Delta\nu^{-1}f_\star^{\rm gal}\frac{\Omega_b}{\Omega_m}M
\nonumber \\
&=6.6\times10^{19}f_\star^{\rm gal} \frac{M}{M_\odot} [{\rm erg~s^{-1}Hz^{-1}}],
\label{LWgal}
\end{align}
where we adopt a mean energy $\bar{E} = 11.7$~eV and bandwidth $\Delta\nu = 5.8\times10^{14}$~Hz for LW photons.
As most of LW photons are produced in 10 Myr, only a fraction 
$\Delta f_{\rm coll}/f_{\rm coll}$ of galaxies are actively producing LW photons, where $f_{\rm coll}$ is the collapse fraction 
in halos above $M_{\rm T4}$, and $\Delta f_{\rm coll}$ the increment of this quantity in the past 10 Myr. 
Similarly to Eq. (\ref{LWgal}), from the {\tt Starburst99} spectrum template we also get the mapping between 
the galaxy luminosity at $1500~\rm\AA$ and the halo mass. Combining these formulae with the halo mass function 
we predict the galaxy luminosity function (LF)  at any redshift. We calibrate the star formation efficiency of normal galaxies, 
$f_\star^{\rm gal}$, by comparing the predicted LF to the observed one 
at $z = 8$ and observed upper limit at $z = 10$ \citep{2011ApJ...737...90B,2012ApJ...745..110O}, finally fixing the star formation efficiency to 
$f_{\star}^{\rm gal} \approx 0.02$.

Some of these normal galaxies might harbor Pop III stars \citep{2011MNRAS.414..847S}, and in principle they should have a slightly different spectrum. However, we do not make such distinction and use the mean LW
luminosity Eq. (\ref{LWgal}) for all star-forming normal galaxies. This appears a reasonable assumption as Pop III stars have short lifetimes, rapidly pollute their hosts,
inducing a transition to Pop II star formation; the galaxy spectrum becomes practically indistinguishable from standard Pop II galaxies \citep{2011ApJ...740...13Z}. 

To gain a quantitative estimate of the relative importance of normal galaxies and DCBHs,
we provide some reference numbers in the following. At $z = 20$, the mean LW luminosity of a star-forming 
normal galaxy in a host with mass $M_{\rm T4}$ is $4.0\times10^{25}~$erg~s$^{-1}$Hz$^{-1}$.
For comparison, the LW luminosity of a $10^6~M_\odot$ DCBH is $2.3\times10^{27}~$erg~s$^{-1}$Hz$^{-1}$. 
Hence, at this epoch, a typical Compton-thick DCBH is more than 50 times brighter than the smallest normal galaxies 
(these small galaxies, however, provide most of the integrated light as they are much more numerous). DCBHs then play a crucial role in prompting further formation of DCBHs in other halos.

Eq. (\ref{dN}) gives the mean number of surrounding halos. To get the probability distribution of the LW specific intensity
as seen by the central halo, we follow \citet{2008MNRAS.391.1961D} and use Monte Carlo simulations. The 
details of the method are given in Appendix \ref{AppA}. By generating a large set of Monte Carlo realizations, we directly get the probability that a central halo sees a 
super-critical LW intensity. We denote this probability by $p_{\rm J}$.

\subsection{Negative feedback}

As already mentioned, for an atomic-cooling halo, being immersed in an intense LW radiation field represents a necessary but not sufficient condition to allow the formation of a DCBH. A second requirement is that the gas of the candidate DCBH host does not contain metals,  which - similar to $\rm H_2$ cooling - would provide the required radiative energy loss to trigger gas fragmentation and star formation. Halos can be enriched by
heavy elements in two independent manners: genetically (i.e. inheriting the heavy elements present in their progenitors) or through winds from neighboring galaxies. In this Section, we study the probability for a halo with $T_{\rm vir} \ge 10^4$~K to be enriched by metals from either its progenitors or neighbors.

\subsubsection{Genetic enrichment mode}\label{genetic}
To study this enrichment mode, we need to understand in detail how metals are propagated through different 
hierarchical halo generations, starting from the very first Pop III stars formed in minihalos. 
 
In the presence of LW radiation, Pop III stars can only form in metal-free minihalos where the gas has enough time to cool and collapse before $\rm H_2$ is dissociated by this radiation. As the cooling and collapse timescales decrease with halo mass, for a given $J_{\rm LW}$ 
Pop III stars can only form in minihalos above a critical mass $M_{\rm crit}^{\rm pop3}$.
This critical mass, as a function of LW intensity, has been obtained via numerical simulations
by \citet{2001ApJ...548..509M} and more recently confirmed by \citet{2008ApJ...673...14O}
and \citet{2007ApJ...671.1559W}.
A widely used fitting formula is
\begin{equation}
M_{\rm crit}^{\rm pop3}( J_{\rm LW}) =1.25\times10^5+2.9\times10^6 J_{\rm LW}^{0.47}~~[M_\odot],
\end{equation}
where $J_{\rm LW}$ is in units of $10^{-21}$erg s$^{-1}$cm$^{-2}$Hz$^{-1}$sr$^{-1}$.
As there is no redshift dependence in the above formula, following \citet{2013MNRAS.432.2909F},
we rewrite it as
\begin{equation}
M_{\rm crit}^{\rm pop3}(J_{\rm LW},z) =M_{\rm crit,0}^{\rm pop3}(z)[1+23.2 J_{\rm LW}^{0.47}],
\label{Mcritpop3}
\end{equation}
where $M_{\rm crit,0}^{\rm pop3}(z)$ is the critical mass in the absence of LW feedback. 
We use $M_{\rm cirt,0}^{\rm pop3}(z) = M_{\rm vir}(z,T_0)$, 
where $T_0$ is the minimum virial temperature above which $\rm H_2$ cooling becomes
efficient. Usually $T_0 \sim(1 - 2)\times10^3$~K \citep{1997ApJ...474....1T}, so we adopt a
value of $2\times10^3$~K in this work. For simplicity, when computing the $M_{\rm crit}^{\rm pop3}$
by Eq. (\ref{Mcritpop3}), we only use the uniform background LW intensity.
This is a conservative assumption as spatial fluctuations from neighboring sources might be
effective in suppressing Pop III star formation in even more massive minihalos, and further 
increase the formation  probability of DCBHs \citep{2012MNRAS.425.2854A}.  

The LW background is the sum of contributions from Pop III stars, star-forming 
normal galaxies and active DCBHs,
\begin{equation}
J_{\rm LW}^{\rm bg}(z) = J_{\rm LW}^{\rm pop3}(z) + J_{\rm LW}^{\rm gal}(z)+ J_{\rm LW}^{\rm BH}(z).  
\end{equation}
We assume that Pop III stars follow a Salpeter IMF in the mass range 1 - 100 $M_\odot$ 
and compute the emission rate of LW photons 
in a Pop III ``instantaneous burst'' mode (more appropriate for the very 
tiny minihalos) according to \citet{2003A&A...397..527S}.  We find that in the ``instantaneous burst" mode most of LW photons are produced in a time interval $\Delta t\approx10^{6.5}$~yr after the burst; the mean photon emission rate is  $\dot{Q}_{\rm LW}\approx10^{46.6}$~photons~s$^{-1}M_\odot$. Repeating the
calculation in Eq. (\ref{LWgal}),
the LW luminosity (per stellar mass) of Pop III stars is then
$l_{\rm LW}^{\rm pop3} = \dot{Q}_{\rm LW}\bar{E}\Delta \nu^{-1} = 1.3\times10^{21}$erg~s$^{-1}$Hz$^{-1}M_\odot^{-1}$.

After the virialization of a minihalo, it will take some more time, $\tau_{\rm d}$, for the gas to cool
and collapse; then Pop III stars form and emit most of their LW photons within $\Delta t$.
The mass density of Pop III stars emitting photons at redshift $z$ is
\begin{equation}
\rho_{\rm pop3}(z) = f_\star^{\rm pop3}\frac{\Omega_b}{\Omega_m}\int_{z}^{\infty}dz^\prime
\int_{M_{\rm crit}^{\rm pop3}(z^\prime)}^{M_{\rm T4}(z^\prime)},
s(M)M\frac{d^2n}{dz^\prime dM}dM
\label{rho_pop3}
\end{equation}
where $f_\star^{\rm pop3}$ is the star formation efficiency of Pop III stars,
$s(M)=1$ for $\tau_{\rm d} < t(z-z') < \tau_{\rm d}+\Delta t$ and equals zero otherwise.

The timescale for gas to cool and collapse is the maximum between the cooling and free-fall timescales 
of a halo,
\begin{equation}
\tau_{\rm d}(M,z^\prime)={\rm max}[\tau_{\rm cool}(M,z^\prime),\tau_{\rm ff}(z^\prime)].
\label{tdelay}
\end{equation}
For a halo with mass $M$ at redshift $z^\prime$,
the $\rm H_2$ cooling timescale for gas enclosed in a radius $r$ is 
\begin{equation}
\tau_{{\rm cool},r}(r,M,z^\prime)=T/\dot{T}=\frac{3k_{\rm B}T_{\rm vir}(M,z^\prime)}{2\Lambda_{\rm H_2}
(\bar{n}_{\rm H}(r),f_{\rm H_2})f_{\rm H_2}},
\end{equation}
where $k_{\rm B}$ is the Boltzmann constant and 
$\bar{n}_{\rm H}(r)$ is the mean number density of the enclosed gas.
We use the cooling function of $\rm H_2$,
$\Lambda_{\rm H_2}$, given by \citet{1998A&A...335..403G},
and always use a maximum $\rm H_2$ fraction of a halo 
$f_{\rm H_2}$ in \citet{1997ApJ...474....1T}, which is independent of $r$.
Then the mean cooling timescale weighted by the mass of gas enclosed within $r$,
\begin{equation}
\tau_{\rm cool}=\left. \int_0^{R_{\rm vir}} \tau_{{\rm cool},r}(r,M,z^\prime)M_{\rm gas}(r)dr
\middle/\int_0^{R_{\rm vir}} M_{\rm gas}(r)dr \right.,
\end{equation}
is used as the cooling timescale of this halo, where $M_{\rm gas}(r)$ is the 
enclosed gas mass derived from the gas profile \citep{1998ApJ...497..555M}.
Analogously, the average of the free-fall timescale weighted by the  
enclosed dark matter mass (Eq. (21) in 
\citealt{2001ApJ...555...92M}) is used as the free-fall time scale of this halo.
We checked that for minihalos with virial temperature $10^3$~K, the $\tau_{\rm d}$
determined above agrees well with the time interval from when the halo reaches 
virial temperature $10^3$~K to the runaway core collapse in simulations 
of \citet{2007MNRAS.378..449G} (panel {\it c} in their Fig. 1 and Fig. 2).

With the mass density of Pop III stars expressed by 
Eq. (\ref{rho_pop3}) we therefore have their contribution to the 
LW background radiation
\begin{equation}
J_{\rm LW}^{\rm pop3}(z) \approx \frac{(1+z)^3}{4\pi} 
l_{\rm LW}^{\rm pop3} \rho_{\rm pop3}(z)l_{\rm max},    
\end{equation}
where
$l_{\rm max}$ is taken to be equal to the mean free path of LW photons set by 
Ly$\beta$ absorption, i.e. the proper distance from $z$ to $z_s=12.1/11.2(1+z)-1$.

The LW radiation contributed by star-forming normal galaxies is
\begin{align}
J_{\rm LW}^{\rm gal}(z)& \approx 
\frac{(1+z)^3}{4\pi}\left[  \int_{M_{\rm T4}}^{M_{\rm 2T4}}(1-f_{\rm BH}) \right.
\bar{L}_{\rm LW}^{\rm gal}(M)\frac{dn}{dM}(z)dM \nonumber \\
&+\left.\int_{M_{\rm 2T4}}^{\infty}\bar{L}_{\rm LW}^{\rm gal}(M)\frac{dn}{dM}(z)dM \right]
\frac{\Delta f_{\rm coll}}{f_{\rm coll}}l_{\rm max};  
\end{align}
and the radiation from active DCBHs is
\begin{align}
&J_{\rm LW}^{\rm BH}(z)\approx 
\frac{(1+z)^3}{4\pi} 
\left[\int_{M_{\rm T4}}^{M_{\rm 2T4}}f_{\rm BH}\bar{L}_{\rm LW}^{\rm BH}
\frac{dn}{dM}(z)dM\right]l_{\rm max},
\end{align}
where $\bar{L}_{\rm LW}^{\rm BH}$ is the mean LW luminosity of active BHs whose
mass distribution is described by Eq. (\ref{dNBH}).
The fraction of active DCBHs, $f_{\rm BH}$, is always defined as 
the number density of active DCBHs
divided by the number density of halos with mass between $M_{\rm T4}$ and $M_{\rm 2T4}$.
Substituting $J_{\rm LW}^{\rm bg}(z)$ for $J_{\rm LW}$ in Eq. (\ref{Mcritpop3}), one can determine 
the critical mass of minihalos for the formation of
Pop III stars, $M_{\rm crit}^{\rm pop3}$, at redshift $z$. 
As $M_{\rm crit}^{\rm pop3}$ at $z$ depends on its value at previous epochs, 
in practice we need to solve  such equation step by step, starting from the redshift 
at which the LW feedback is negligible, i.e., 
the critical mass is $M_{\rm crit,0}^{\rm pop3}(z) = M_{\rm vir}(z,T_0)$. 

Having obtained $M_{\rm crit}^{\rm pop3}$, for each halo we can calculate the probability that its most massive progenitors (the earliest progenitors with a certain mass) ever hosted Pop III stars. The maximum of these probabilities for all possible masses of the most massive progenitors is the probability of genetic enrichment \citep{2009ApJ...694..879T,2007ApJ...667...38T}.
To get this probability, the distribution of the formation redshift of the 
most massive progenitors is needed.  For a halo with mass $M_0$ at redshift $z_0$, the probability that 
its most massive progenitor with $M$ formed before redshift $z$ is (fitted by \citealt{2012MNRAS.422..185G} according to their simulations),
\begin{equation}
P(>z,M|M_0,z_0)=\frac{\alpha_f}{{\rm exp}(w^2/2)+\alpha_f-1}
\label{Pf}
\end{equation}
where 
\begin{equation}
\alpha_f=0.867{\rm exp}(-2f^3)/f^{0.8},
\end{equation}
$f=M/M_0$
and
\begin{equation}
w=\frac{\delta_c(z)-\delta_c(z_0)}{\sqrt{\sigma^2(fM_0)-\sigma^2(M_0)}}.
\end{equation}
Finally we get the genetic probability via the following steps. For a progenitor of mass $M$, using Eq. (\ref{tdelay}), we determine 
a redshift $z_{\rm crit}$ defined as the maximum of $z+\Delta z_{\rm d}$ and the redshift when $M_{\rm crit}^{\rm pop3}(z^\prime) = M$,
where $\Delta z_{\rm d}(M,z^\prime)$ is the redshift interval corresponding to a time $\tau_{\rm d}(M,z^\prime)$ before $z$. 
Progenitor with mass $M$ must have formed before $z_{\rm crit}$, otherwise Pop III stars would fail to form before $z_0$ in halos with mass $M_0$ (no enough time for the gas to cool and collapse).
The formation probability of Pop III stars in progenitor of $M$ 
before $z_{\rm crit}$ is therefore $P(>z_{\rm crit},M|M_0,z_0)$. The final genetic enrichment probability of a host halo $M_0$ at redshift $z_0$, $p_{\rm g}(M_0,z_0)$, is the maximum probability obtained after scanning all possible $M< M_0$ values. 

\subsubsection{Wind enrichment mode}

The potential birth place of a DCBH might also be polluted by metals coming 
from neighboring galaxies. 
Metals can be carried to large distances from their origin halos by galactic winds expanding at a speed $v_{\rm W}$.
Hence, the enrichment probability in the wind mode, $p_{\rm W}$, is just the probability for a halo that has neighboring 
galaxies within the radius $v_{\rm W}\times t(z_{\rm F},z)$ (due to the finite energy of supernovae, there is an upper limit 
$r_{\rm W}\sim 0.2$ comoving Mpc on the propagation distance of the wind; $v_{\rm W} \sim 50$ km/s for normal galaxies, see Fig. 1 and Fig. 2 in \citealt{2003ApJ...588...18F}), where $z_{\rm F}$ is the redshift at which the wind is first ignited. 
Taking $z_{\rm F}$ as the redshift at which the most massive progenitor of a halo reaches a mass\footnote{We assume that metals are ejected by all halos above $M_{\rm T4}$, as long as they do not contain {\it active} DCBHs. Even if we have assumed that only a fraction $\Delta f_{\rm coll}/f_{\rm coll}$ of them are star-forming, the remaining halos, including those that previously had active DCBHs, 
must have experienced star formation after the BH stop accreting gas,
see \citet{2013MNRAS.432.3438A}.} of $M_{\rm T4}$, the probability distribution of $z_{\rm F}$ is again calculated using 
Eq. (\ref{Pf}). The computation therefore is very similar to the $p_{\rm J}$ and we carry it on together with that using the same Monte Carlo simulations.

In addition to normal galaxies, minihalos may also eject metals into the environment if they experienced Pop  
III star formation. However, in this case the propagation distance of metals is small and 
if the minihalos are massive metals even cannot escape the host halos \citep{2008ApJ...682...49W}.
In addition, \citet{2008ApJ...674..644C} pointed out that the mixture of metals carried by galactic winds
with gas in halos  is not effective enough, while \citet{2013ApJ...773...19M} found that
the metals are re-accreted by the host halos and cannot significantly enrich the environment.
We therefore ignore this contribution to the metal enrichment.

We only consider the influence of metal enrichment on central halos with $T_{\rm vir} \ge 10^4$~K, as the major mechanism 
quenching Pop III star formation in minihalos is LW feedback, fully captured by $M_{\rm crit}^{\rm pop3}$. 
We believe that this is a good approximation at $z$ \simgt 25, considering that the formation of Pop III stars is quenched quickly by DCBHs shortly thereafter according to our results.

\subsubsection{Photoevaporation feedback}\label{photoevaporation}

Photoevaporation plays a significant role in suppressing  galaxy/DCBH formation, particularly in smaller 
halos that could be fully photoevaporated by ionizing radiation.
For halos with mass $M_0$ at $z_0$, at any $z > z_0$, the median mass of their
most massive progenitors is obtained by 
taking $P = 0.5$ in Eq. (\ref{Pf}) \citep{2012MNRAS.422..185G}. We use the evolution of the median
of the most massive progenitors as a typical growth history of 
a halo with mass $M_0$ at $z_0$.
The evolution of the LW intensity as seen by this typical most massive progenitor is expressed by
Eq. (\ref{JLW}) using $N_{\rm P} = N\Delta M_i \Delta l _j$ in each cell, and replacing $m$ with this typical progenitor 
mass at each redshift.  The ionization intensity (at 13.6 eV) as seen by a halo located in an ionized bubble
is derived from the LW intensity of normal galaxies alone, as DCBHs are highly obscured at energies 
$>$ 13.6 eV, and photoevaporation becomes significant only at later redshifts ($<\sim20$)
when the formation of Pop III stars is almost totally quenched in our models. If a halo resides outside ionized bubbles, 
the ionizing radiation intensity vanishes, since the optical depth  of neutral gas to ionizing photons is very high.
Therefore, the ionizing intensity can be expressed as

\begin{equation}
J_{\rm ion}=\left\{
\begin{tabular}{ll}
$f_{\rm ion}J_{\rm LW}^{\rm gal}/\delta$~~~&when $J_{\rm LW}^{\rm gal} > J_{\rm LW}^{\rm eq}$\\ 
0&when $J_{\rm LW}^{\rm gal} < J_{\rm LW}^{\rm eq}$,
\label{JionJLW}
\end{tabular}
\right.
\end{equation}
where $f_{\rm ion} \le 1$ is a normalization factor considering the combination of 
escape fraction and the absorption due to neutral absorbers in the IGM.
For normal galaxies, the spectral ratio $\delta\approx2.8$.
$J_{\rm LW}^{\rm eq}$ is the LW intensity corresponding to an ionizing 
field intensity set by ionization equilibrium: seeing an ionizing intensity
larger than $J_{\rm LW}^{\rm eq}$ indicates that the halo must reside in an ionized bubble. 

If the ionizing radiation has a $\nu^{-5}$ power-law shape \citep{2013MNRAS.432L..51S}, considering that the ionization cross-section has instead a $\nu^{-3}$ form, the ionization intensity at ionization equilibrium is
$J_{\rm ion}^{\rm eq}=8n_{\rm H,0}(1+z)^3\alpha_Bh_px^2_{\rm H}/[4\pi(1-x_{\rm H})\sigma_0]$,
where $n_{\rm H,0}$ is the comoving hydrogen number density, $\alpha_B$ is the case B recombination rate, $h_p$ is the Planck constant, $\sigma_0$ is the H-ionization cross-section at 13.6 eV, $x_{\rm H}$ is the neutral hydrogen fraction at the edge of
 ionized bubbles for which we adopt a value $x_{\rm H}= 0.001$.

Let us define the redshift at which a given halo is first engulfed by the expanding ionization front and embedded into the 
bubble as $z_{\rm IN}$: this is the moment at which photoevaporation of the halo gas starts. We compute the fraction of 
the original baryons left in $M_0$ at $z_0$ using Eq. (6) in \citet{2013MNRAS.432L..51S}. Formulae in \citet{2013MNRAS.432L..51S} 
are fitted from results of simulations assuming  constant ionization intensity; however, in our calculation, from $z_{\rm IN}$ to $z_0$, we find that the ionization intensity seen by the typical most massive progenitor increases by about two orders of magnitude, see the upper panel of Fig. \ref{Jion}.
Therefore we always use the $J_{\rm ion}$ seen by this progenitor at a median 
redshift $(z_0+z_{\rm IN})/2$ as an effective intensity. We assume the DCBH could only form in halos whose mass is above $M_{\rm T4}$, and have 
at least 20\% leftover gas content. The net result of photoevaporation feedback is to boost the minimum mass, 
$M_{\rm min}^{\rm ion}$, of galaxy or DCBH host halos. 

\subsection{DCBH abundance}

At redshift $z$, the number density of halos with mass between $M$ and $M+dM$ is $\frac{dn}{dM}(z)dM$; among these halos, a fraction $\frac{dP}{dz^\prime}(z^\prime,M^\prime_{\rm T4}|M,z)dz^\prime$ ($P$ is given by Eq. (\ref{Pf})) have most massive progenitors with mass $M^\prime_{\rm T4}$ -  $M_{\rm T4}$ at redshift
$z^\prime$ - between $z^\prime$ and $z^\prime+dz^\prime$.
The time interval between $z^\prime$ and $z$ is that required for the gas to collapse into a BH. 
Such timescale has been estimated to be of several Myr \citep{2013ApJ...774..149C},
therefore we use 5 Myr here. Some of these progenitors have already been polluted or were not exposed to a strong enough LW radiation, therefore BHs 
only form in a fraction
$[1-p_{\rm g}(M^\prime_{\rm T4},z^\prime)]p_{\rm J-W}(M^\prime_{\rm T4},z^\prime)$ 
of them, where $p_{\rm J-W}$ is the probability for a halo exposed to a super-critical
LW radiation and not
polluted by metals from neighboring galaxies,
a quantity directly obtained from our Monte Carlo simulations.
We solve for
$p_g$ analytically
as described in Sec. \ref{genetic}.
The formation rate of DCBHs at $z$ is then
\begin{align}
\frac{dn_{\rm BH}}{dz}(z)&=[1-p_{\rm g}(M^\prime_{\rm T4},z^\prime)]p_{\rm J-W}(M^\prime_{\rm T4},z^\prime) \nonumber \\
&\times \int_{M_{\rm T4}}^{M_{\rm 2T4}} dM\frac{dn}{dM}(z)\frac{dP}{dz^\prime}(z^\prime,M^\prime_{\rm T4}|M,z)\frac{dz^\prime}{dz}.
\label{dnBHdz}
\end{align}

With Eq. (\ref{dnBHdz}), the number density of {\it active} DCBHs at 
$z$ is then
\begin{equation}
n_{\rm BH}(z)=\int_z^{z_{t}}\frac{dn_{\rm BH}}{dz^\prime}dz^\prime,
\label{nBH}
\end{equation}
where $z_{t}$ is the redshift 
corresponding to a time $t_{\rm QSO}$ before $z$ (DCBHs formed earlier than $z_{t}$ have already exhausted/ejected their
gaseous fuel and therefore faded away at $z$).

The combination of Eqs. (\ref{dnBHdz}) and  (\ref{nBH}) is an integro-differential equation and is integrated step by step.
We start from a high redshift, e.g. 50, when the number of DCBHs is negligible, so the initial $n_{\rm BH}$ is set to be zero.
After each step of the integration, we update $n_{\rm BH}$. Note both $M_{\rm crit}^{\rm pop3}$ and $M_{\rm min}^{\rm ion}$
depend on the evolution history of $n_{\rm BH}$ before this redshift, so we also update them accordingly. We then update $p_{\rm g}$ by using the new $M_{\rm crit}^{\rm pop3}$, and $p_{\rm J-W}$ by using the new $n_{\rm BH}$ in the Monte Carlo realizations.
We then finally get the evolution of the number density of DCBHs.

The formation rate in Eq. (\ref{dnBHdz}) also allows to derive the mass density of active DCBHs:
\begin{equation}
\rho_{\rm BH}(z)=\int_z^{z_{t}}M_{\rm BH,seed}
{\rm exp}(t^\prime/t_{\rm Edd})\frac{dn_{\rm BH}}{dz^\prime}dz^\prime,
\label{rhoBH}
\end{equation}
where $t^\prime$ is the time interval between
$z_{t}$ and $z^\prime$,
and the cumulative (i.e. active + dead) BH mass density 
\begin{equation}
\rho_{\rm BH}^{\rm cum}(z)=\int_z^{\infty}M_{\rm BH,seed} e^{[{\rm min}(t^\prime,t_{\rm QSO})/t_{\rm Edd}]}
\frac{dn_{\rm BH}}{dz^\prime}dz^\prime.	
\end{equation}

\section{RESULTS}\label{results}
\begin{table}
\begin{center}
\begin{tabular}{lccccccc}
\hline \hline
&${\rm model} $ & $J_{\rm LW}^{\rm crit}$  &  $l_{\rm min}$ &$f_{\rm ion}$ &genetic enrichment \\
\hline
&$A$ &50&   1 & 0.2&Yes \\
&$B$ &150&  1 & 0.3&No  \\
&$C$ &30&   2 & 0.3&Yes \\
\hline
\hline
\end{tabular}
\caption{Parameters in different models, 
$J_{\rm LW}^{\rm crit}$ is in units of 
$10^{-21}$erg~s$^{-1}$cm$^{-2}$Hz$^{-1}$sr$^{-1}$,
while $l_{\rm min}$ is the minimum distance between
two halos in units 
of the sum of the virial radii of them.
Parameters not listed in 
this table are the same for all models, 
e.g., $T_0 = 2\times10^3$~K, $f_{\star}^{\rm pop3} = 10^{-3}$, 
$v_{\rm W} = 50$~km/s and $r_{\rm W} = 0.2$~comoving Mpc.
Model $A$ is our fiducial model.} 
\label{table:parameters}
\end{center}
\end{table}
									
\begin{figure}
\begin{center}
\subfigure{\includegraphics[scale=0.4]{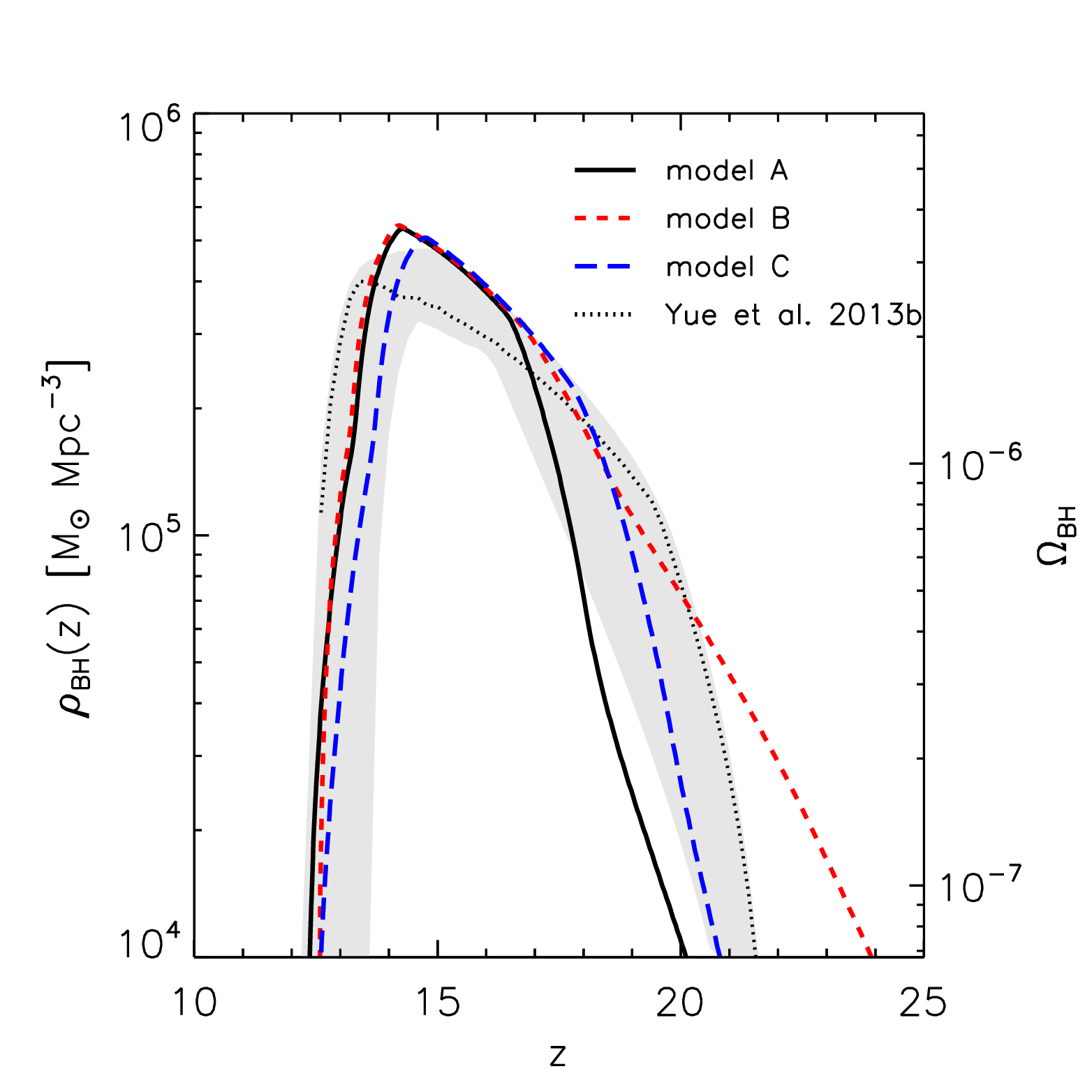}} 
\subfigure{\includegraphics[scale=0.4]{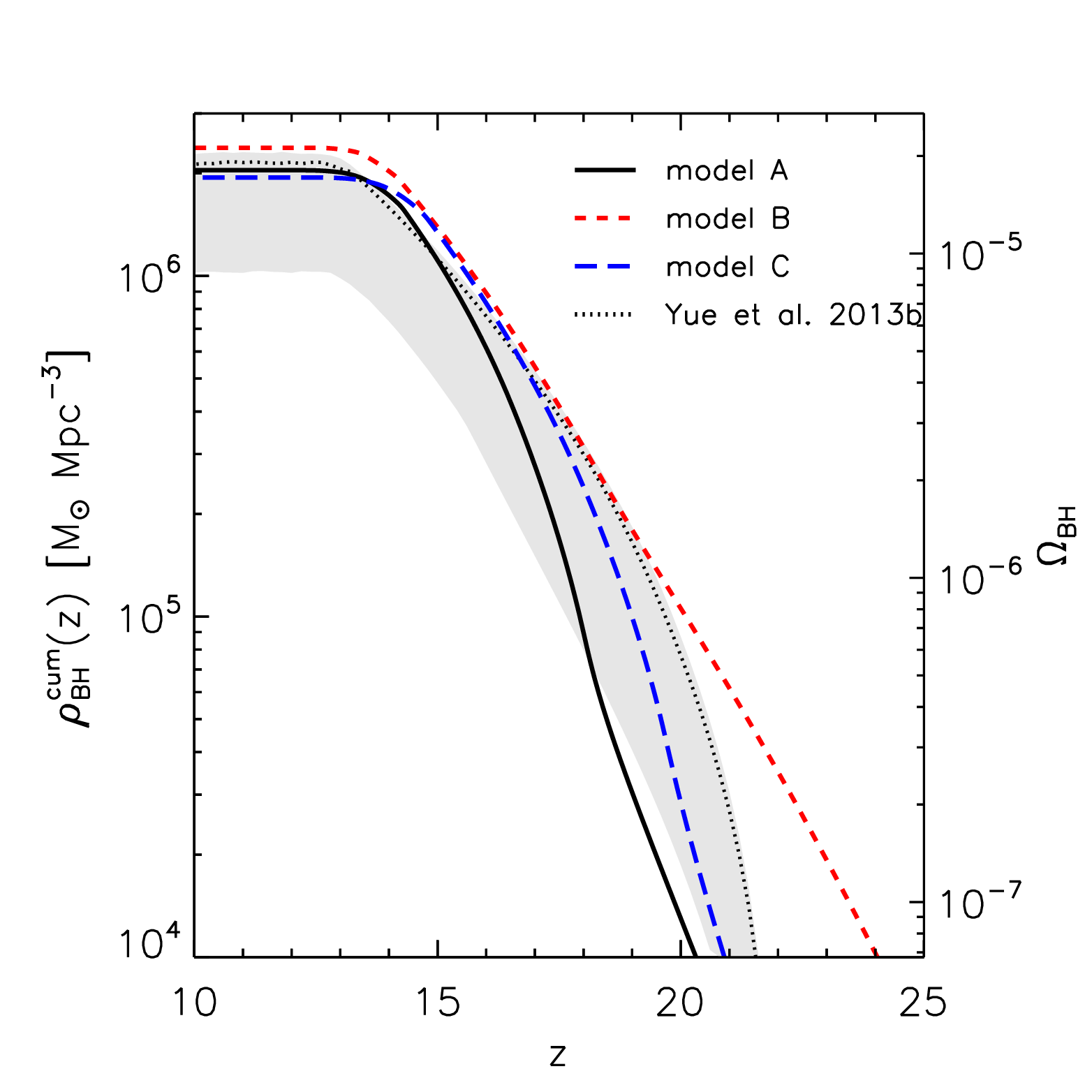}}
\caption{\textit{Left:} Evolution of the mass density of DCBHs 
from models $A$, $B$ and $C$ (see Tab. \ref{table:parameters}) 
of this paper, 
and the mass density expected in \citet{2013MNRAS.433.1556Y} 
(the thin dotted line) with 1$\sigma$ uncertainties of the parameters
(shaded).
\textit{Right:} same for the cumulative mass density.}
\label{rho_BHz}
\end{center}
\end{figure}

\begin{figure}
\begin{center}
\includegraphics[scale=0.4]{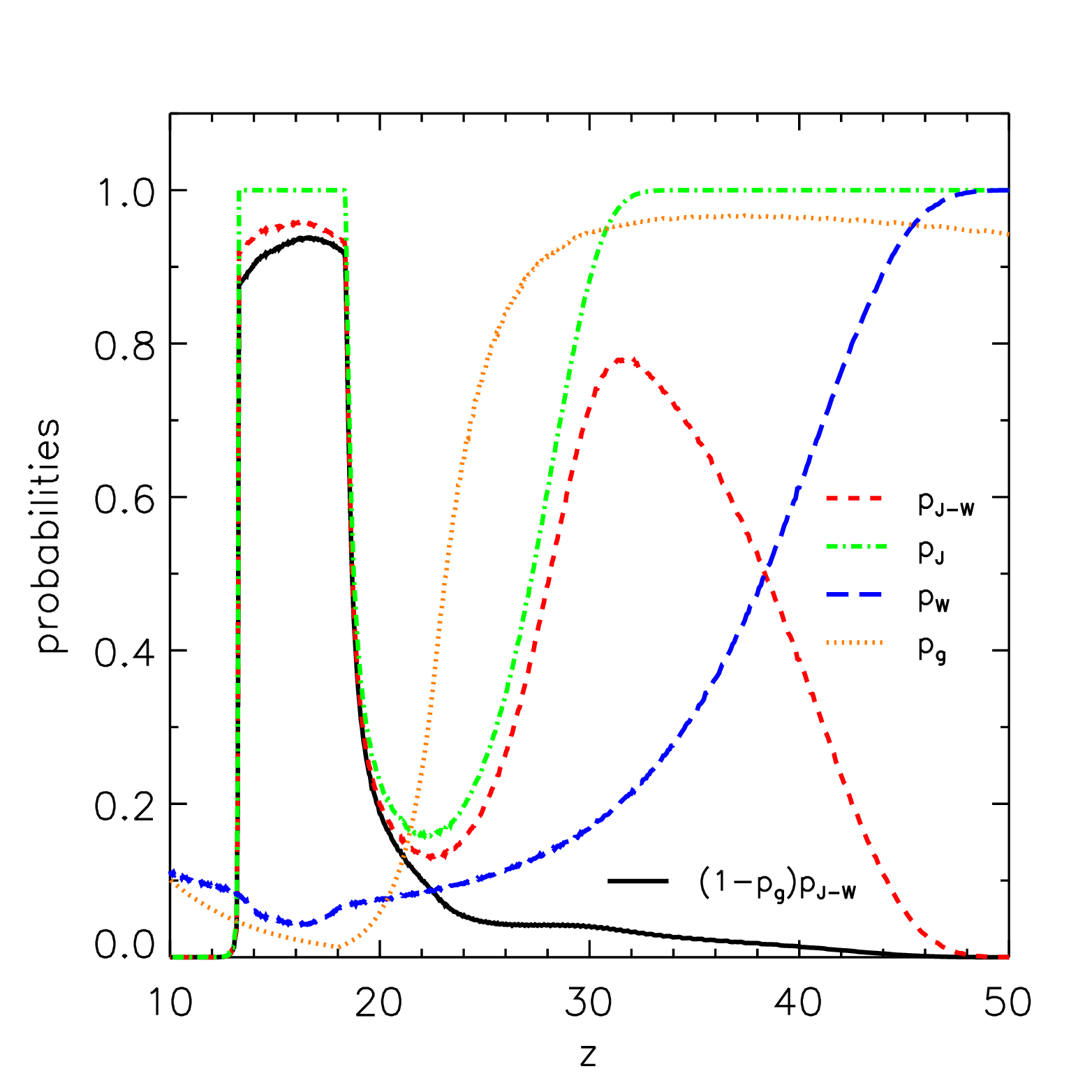}
\caption{Redshift evolution of the probabilities for the different processes considered (see text) in the fiducial model $A$.}
\label{probabilities}
\end{center}
\end{figure}

In Fig. \ref{rho_BHz}, we plot the evolution of the mass density (left panel) of active DCBHs, and the black hole cumulative mass density (right) for our models, whose parameters are listed in Tab. \ref{table:parameters}. 
In each panel, as a reference we add the best-fitting mass density evolution 
required to fit NIRB observations \citep{2013MNRAS.433.1556Y}, and the mass density range corresponding to $1\sigma$ uncertainties of the model free parameters in that paper. In all models, the adopted $J_{\rm LW}^{\rm crit} $ 
is consistent with the critical intensity given by \citet{2010MNRAS.402.1249S} for a $10^4$~K black-body spectrum,
which is between 30 and 300. To gain further insight into the prediction of the model, we plot in Fig. \ref{probabilities} the evolution of $p_{\rm J-W}$, $p_{\rm J}$, $p_{\rm W}$, $p_{\rm g}$ and $(1-p_{\rm g})p_{\rm J-W}$ for a halo with mass $M_{\rm T4}$ in the fiducial model. We recall that the $(1-p_{\rm g})p_{\rm J-W}$ describes the formation probability of the DCBH in recently formed $M_{\rm T4}$ halos. 

\subsection{Fiducial model}
Let us start from the analysis of the fiducial model results. Initially, all halos with mass $M_{\rm T4}$ are 
exposed to a LW intensity above $J_{\rm LW}^{\rm crit}$, as at high redshift they represent rare, highly biased  
peaks of the density field (dash-dotted line in Fig. \ref{probabilities}). However, DCBHs can hardly form in such systems
as the neighboring galaxies providing the necessary LW illumination also eject metals, yielding $p_{\rm W} \simeq 1$ 
and $p_{\rm J-W} \simeq 0$ (long and short-dashed lines, respectively). During cosmic evolution, these halos become 
less clustered and the metal enrichment probability by winds decreases considerably. As a result, at $z\sim 45$  
$p_{\rm W}$ starts to drop and $p_{\rm J-W}$ starts to rise. At this time some DCBHs start to form in halos containing pristine gas;
once formed, they provide extra LW photons in addition to normal galaxies.
As the universe keeps expanding, eventually $p_{\rm J}$ also starts to drop ($z\sim30$) at 
a high rate, 
because at this time the previously formed DCBHs are still too few to provide a sufficiently large number of LW photons.
As a result, the $p_{\rm J-W}$ reaches a peak of $\sim0.8$ at $z\sim30$, then it drops again, reaching the trough $\sim0.15$ 
at $z\sim22$. The combined probability $(1-p_{\rm g})p_{\rm J-W}$ closely traces the $p_{\rm J-W}$ trend albeit with a 
much smaller value as the majority of halos ( \simgt 90\%) are polluted by metals from their progenitors (solid line). 
Note that since $z\sim45$ and until
$z\sim20$, the genetic enrichment is the dominant 
metal enrichment (and DCBH suppression) mode. 

A simple analogy will help understanding the physical evolution at this point. A DCBH can be visioned as a flame, with the newly formed pristine gas halos representing dry wood. If enough dry wood is provided the flame will grow stronger and dry more other wood by itself, reaching a kind of self-supported regime; alternatively, the flame will rapidly extinguish due to the lack of fuel supply. The period $z\sim 30 -20$ represents a dangerous stage for the formation of DCBHs, as 
initially a large fraction of $M>M_{\rm T4}$ halos is already polluted, leaving only a minority of pristine ones 
as candidate hosts of new DCBHs. If the drop of $p_{\rm J}$ would continue beyond that epoch, the initial DCBH ``weak flame'' would 
be extinguished or remain very weak. This would be the result if the LW radiation from DCBHs themselves is not considered. 

DCBHs are about two orders of magnitude brighter than most of normal galaxies in the LW band. Thus, even if they are rare at 
high redshifts,  they dramatically impact cosmic structure formation in several ways. The first obvious effect is that the formation probability of Pop III stars in minihalos is reduced significantly and even quenched after $z\sim 21$. Therefore the genetic 
enrichment probability is reduced (i.e. minihalos are sterilized) and the fraction of $M_{\rm T4}$ halos with 
pristine gas increases. Secondly, LW radiation from previously formed DCBHs can trigger the formation of further DCBHs
much more efficiently than normal galaxies. For example, at $z = 30$, the LW intensity of a normal galaxy in a host halo with mass
$M_{\rm T4}$ drops to 30 at a distance $l\approx0.7$~kpc (physical). The distance is only slightly larger than $2R {\rm vir}$ of the host halo, so 
it is only marginally possible for 
a single normal galaxy to trigger the formation of DCBHs at this redshift. On the other hand, for a BH with mass of $10^6~M_\odot$, the corresponding distance is about 7 kpc. At the same redshift, in the volume within this distance, during the accretion timescale of a BH, about 50 new halos with mass between $M_{\rm T4}$ and $M_{\rm 2T4}$ form. It means that a single BH can ideally prompt the formation of about 50 new DCBHs! In practice, though, this number is likely smaller because at such high redshift many of these halos are metal-enriched when they form. Whatever its precise strength is, this positive feedback exerted by 
DCBHs on their own formation can hardly be overlooked.

The final results depend on the competition between the negative factors (the expansion of the universe and metal pollution) and the above positive feedback. If the latter dominates, the formation probability of the DCBH will rise again. At $z \sim 22$, the abundance of DCBHs increases to a level that the trend of decreasing $p_{\rm J}$ set by the negative factors can be reversed, $p_{\rm J}$ and $p_{\rm J-W}$ then start to rise, while around this time $p_{\rm g}$ drops very fast. The rise of the DCBH population and the sharp decrease of the genetic enrichment are interlinked, of course. The universe then at $z\sim 20$ enters the DCBH era, in which 
$p_{\rm J}$ grows so rapidly due to positive feedback that essentially all newly formed, unpolluted $M_{\rm T4}$ halos become 
populated with active DCBHs, whose density reaches a peak of about $5\times10^5~M_\odot$Mpc$^{-3}$
at $z\sim14$ (model $A$ in Fig. \ref{rho_BHz}, left panel).  The DCBH era extends until $z\sim 13$, or about 150 Myr.

\begin{figure}
\begin{center}
\includegraphics[scale=0.4]{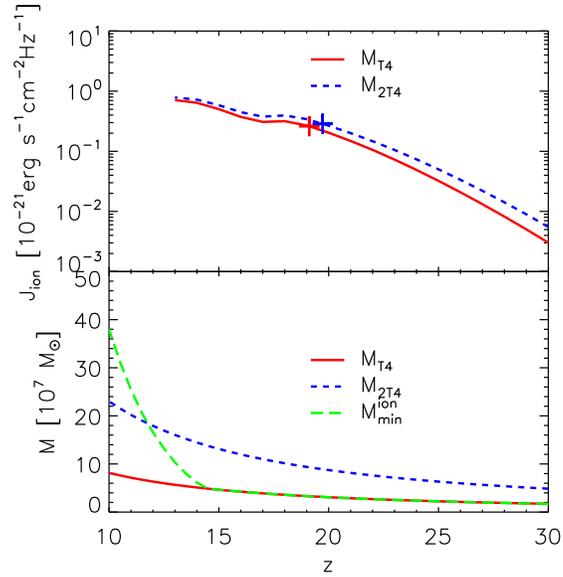}
\caption{{\it Upper:} Ionizing background intensity seen by the typical progenitor of a halo with mass 
$M_{\rm T4}$ (solid) and $M_{\rm 2T4}$ (dashed) at redshift 13. We mark $z_{\rm IN}$ by crosses. {\it Bottom:} Critical mass of halos 
that can retain at least 20\% of their gas against photoevaporation (long-dashed), $M_{\rm T4}$ (solid) and $M_{\rm 2T4}$ (dashed).}
\label{Jion}
\end{center}
\end{figure}

What causes the end of the DCBH era? The answer is \textit{photoevaporation}, which represents a third negative feedback - in addition to cosmic expansion and metal enrichment - that sets in at later stages. To understand this key point let us inspect the  
upper panel of Fig. \ref{Jion}, where we show the ionizing background intensity, $J_{\rm ion}$,
as seen by the most massive progenitors of two halos of mass $M_0 = M_{\rm T4}$ and $M_0 = M_{\rm 2T4}$ at $z_0 = 13$, respectively. We mark $z_{\rm IN}$  
on each curve. We also plot (bottom panel) the minimum mass of halos in which DCBHs could still form even under the effects of photoevaporation. Photoevaporation feedback does not affect the formation of 
DCBHs before $z\sim 14$, when even halos with mass $M_{\rm T4}$ are massive enough to hold more than 
20\% of their gas (long-dashed curve in the bottom panel of Fig. \ref{Jion}). However, after then DCBHs could only form in halos with $M>M_{\rm min}^{\rm ion}$, which increases steadily since $z\sim 14$.
For example, at $z\sim 13$ $M_{\rm min}^{\rm ion}$ is about 1.8 times $M_{\rm T4}$. The number of newly formed halos with 
$M<M_{\rm 2T4}$ which can hold at least 20\% of their gas is reduced by about 70\%. 
As photoevaporation feedback becomes effective, the abundance of DCBHs decreases (after a time delay $\sim t_{\rm QSO}$) steadily until a sharp drop occurs at $z\sim 13$, when even halos as massive as $M_{\rm 2T4}$ cannot prevent their gas from being almost completely photoevaporated.

We stress that the above scenario is largely independent of the upper mass limit of DCBH host halos. 
Adopting a higher mass cut would make the DCBH disappearance phase slightly more gentle as a few among
the most massive host halos are able to retain their gas. Considering also the uncertainties in the details of our 
photoevaporation model, it is clear that the detailed late evolution of the DCBH abundance deserves
additional future work.

\subsection{Model variations}\label{modvar}

Besides the fiducial one, we have explored different models (see Tab. \ref{table:parameters}) 
to check the robustness of our conclusions. 
The evolution of mass density in these models is also plotted in Fig. \ref{rho_BHz}.

The first variation is model $B$ without genetic enrichment. This is motivated by the possible suppression of Pop III star formation in 
progenitor minihalos exposed to a sufficiently large LW intensity $\sim 10^3\times10^{-21}$~erg s$^{-1}$cm$^{-2}$Hz$^{-1}$sr$^{-1}$, as proposed by \citet{2012MNRAS.422.1690P}. This is combined with a larger $J_{\rm LW}^{\rm crit}=150$ value. We confirm the rise of the DCBH population, which now occurs earlier and reaches a higher peak  $\approx 6\times10^5$~$M_\odot$Mpc$^{-3}$ (model $B$ in Fig. \ref{rho_BHz}).

With model $C$ in Fig. \ref{rho_BHz} we check the sensitivity of the results to the assumed
minimum interhalo distance. As shown by Fig. \ref{cfl}, at redshift 30, if the sources of the radiation received
by a central $M_{\rm T4}$ halo are halos with mass $> M_{\rm T4}$,  $>90$\% of the radiation
comes from neighbors within 10$R_{\rm vir}$. However, close neighbors may merge with the central halo 
on a short timescale. To assess the impact of possible mergers on our conclusions, model $C$ ignores the 
contribution from close neighbors within a certain distance by adopting a larger $l_{\rm min}$, and uses the 
minimum $J_{\rm LW}^{\rm crit}$  given in \citet{2010MNRAS.402.1249S}. We posit $l_{\rm min }$ equal to
2 times the sum of the virial radii of two halos.

The basic change with respect to the fiducial model is that both $p_{\rm J}$ and $p_{\rm W}$ are reduced by the larger $l_{\rm min}$. However, as metal enrichment by galactic winds is more sensitive to the distance between halos, at early stages $p_{\rm J-W}$ is somewhat higher than in the fiducial model. 
An even larger $l_{\rm min}$ value would fail to produce the DCBH population rise after $z \sim 20$;
if genetic enrichment is turned off, though, the rise can occur even for  $l_{\rm min}=5$.

\section{CONCLUSIONS}\label{conclusions}

We have studied the formation and evolution of DCBHs at high ($z$ \simgt 13) redshifts. 
The most striking result is that we found that a well-defined DCBH     
formation era might occur in the early universe extending approximately for 150 Myr 
from $z\sim 20$ to $z\sim 13$.
The intermediate masses ($M\sim10^{6} M_\odot$) of DCBHs make them excellent seeds for the subsequent
rapid growth of the SMBH observed at $z$ \simgt 6.
In addition, such BHs could be very abundant in this era and might 
leave a clear signature in the observed fluctuations of the source-subtracted NIRB
\citep{2013MNRAS.433.1556Y}.

During such era the abundance of active DCBH grows, reaches a peak and eventually comes to a sudden halt when the required conditions for their formation disappear. These conditions include the availability of pristine gas in atomic-cooling halo ($T_{\rm vir} \simgt 10^4$~K) embedded in a sufficiently strong LW
or NIR radiation field. Whether and for how long these conditions can be met during cosmic evolution depends on a complex network of negative and positive feedbacks that we have described in detail.  

A key result is that the (so far neglected) positive feedback of DCBH on their own formation fosters a rapid multiplication of the number of DCBHs. This is possible because DCBHs 
are likely 
Compton-thick systems (see \citealt{2013MNRAS.433.1556Y}), they are very bright in the rest-frame UV/NIR band,
typically outshining small high-redshift galaxies by more than 10 times; in their surroundings the LW flux is therefore largely enhanced with respect to the level provided by galaxies, thus enabling additional DCBHs to form. DCBH formation spreads very rapidly, similarly to a flame in a haystack. Their mass density raises from $\sim 5$~$M_\odot$Mpc$^{-3}$ at $z \sim 30$ to the peak value $\sim5\times10^5$~$M_\odot$Mpc$^{-3}$ (corresponding to a density parameter $\Omega_{\rm BH} \sim 3\times10^{-6}$)  at $z \sim 14$ in our fiducial model.

The DCBH formation era however does not last very long. This is due to a combination of several negative feedback effects that we have discussed in Sec. \ref{methods}. In addition to the effect of a decreasing bias of DCBH host halos with time, metal pollution, either due to heavy elements inherited by the progenitor halos or carried by supernova winds coming from neighbors, is perhaps an obvious way to quench DCBH formation. However, we find that collectively these effects are not able to quench the DCBH 
widespread formation before $z\sim6 - 7$. Instead, the rapid DCBH birthrate drop is caused by photoevaporation of the gas from the
gravitational well of the smallest atomic-cooling halos.  This effect reduces the 
formation rate of the DCBH, almost completely suppressing their formation for  $z\simlt 13$. This is the reason why DCBH formation era lasts for a relatively short time interval during cosmic evolution. However, as they are Compton-thick, their ionizing photons are fully reprocessed into lower energy radiation before escaping into the IGM and, as such, they play no active role in driving reionization.   

The above scenario is robust against variations of the free parameters of the model,
for which we have explored (see Sec. \ref{modvar}), including the critical LW intensity threshold for DCBH formation, the strength of the genetic enrichment, and the influence of the assumed minimum distance between two halos. In spite of these, we do not observe substantial 
deviations from the evolutionary trend described by the fiducial case, witnessing a DCBH formation history firmly controlled by feedback(s) action.

Our work is based on the most popular assumption that DCBH can only form under specific conditions,
i.e., the gas is metal-free and the $\rm H_2$ formation is suppressed 
by strong external radiation. It is possible that  
even in the presence of metals or $\rm H_2$, if the accretion timescale is smaller than the 
fragmentation timescale, DCBH may still form  \citep{2009ApJ...702L...5B,
2010Natur.466.1082M,2014MNRAS.437.1576B}; particularly because the supersonic turbulence
suppresses the fragmentation \citep{2005ApJ...630..250K,2007ApJ...656..959K}. 
Albeit interesting, this scenario still needs to overcome a number of open questions that have been discussed 
in a recently published paper by two of the present authors \citep{2013MNRAS.434.2600F}.
We cannot exclude different DCBH formation channels with respect to the canonical one explored here,
however, the photoevaporation of haloes with circular velocity \simlt 30~km/s, 
which is discussed in this paper, might in practice make these channels only have modest 
influence on our conclusions. We also note that purely forming DCBH in metal free, 
UV illuminated halos already produces a BH cosmic mass density comparable to the present-day one.

We finally comment on a possibly intriguing implication of our model. The end of the DCBH era leaves behind a large number of ``fossil'' DCBH seeds
($\Omega_{\rm BH} \sim 10^{-5}$). 
Therefore to have sufficient number of seeds for 
SMBHs is not a problem any more in this scenario
(see also \citealt{2012MNRAS.425.2854A}).
Some of DCBHs are ready to be embedded in other 
galaxies and in SMBHs. While only a small fraction of these BHs will continue to grow following merging events and possibly becoming SMBHs, the majority of them will evolve passively. 
Hence a large number of intermediate-mass black holes could be present in local universe, which has escaped so far detection. In the future it will be  interesting to consider suitable strategies to find them. It is also tempting to speculate that DCBH gas--devoid, low-mass dark matter halo hosts might be related with the large missing halo population predicted by $\Lambda$CDM models at $z=0$.

\section*{ACKNOWLEDGMENTS}
It is a pleasure to acknowledge M. Dijkstra, S. Salvadori and E. Sobacchi for discussions and comments.
We acknowledge financial support from PRIN MIUR 2010-2011, 
project ``The Chemical and Dynamical Evolution of the Milky Way and Local Group Galaxies'', prot. 2010LY5N2T.
BY and XC also acknowledges the support   
of the NSFC grant 11073024 and the MoST Project 863 grant 2012AA121701. YX is supported by China             
Postdoctoral Science Foundation and by the Young Researcher Grant of National Astronomical Observatories,    
Chinese Academy of Sciences.


\appendix

\section{LW field realizations}\label{AppA}
Eq. (\ref{dN}) gives the mean number of surrounding halos. 
To get the probability distribution of the LW specific intensity
as seen by the central halo, we follow \citet{2008MNRAS.391.1961D},
we generate a large number of Monte Carlo realizations by dividing the mass and distance ranges of the surrounding halos into $N_{M}\times N_{l}$ bins. In the $(i,j)$-th bin with central mass and central distance ($M_i,l_j$), there are $N(m,z,M,l)\Delta M_i \Delta l_j$ halos on average.
The actual number in each bin for each realization, $N_{\rm P}$, follows the Poisson distribution with this mean. Regarding to the minimum distance between two halos, $l_{\rm min}$, 
a natural choice is the sum of their virial radii, but we also discuss models with higher minimum distance.
The maximum distance is set by Ly$\beta$ absorption,
as described in Sec. \ref{genetic}.
In each realization, the LW specific intensity as seen by the central halo is\footnote{
Cosmological corrections are not considered, but it is safe enough for our case \citep{2008MNRAS.391.1961D}.
In principle the absorption due to the residual $\rm H_2$ in the IGM 
should be included, however, this effect is rather uncertain and complicated, 
we therefore have to ignore it in current work.}
\begin{equation}
 J_{\rm LW}=\frac{1}{4\pi}\sum_{i=1,j=1}^{N_{M},N_{l}}\frac{\sum_{k=1}^{N_{\rm P}}L_{\rm LW}^{k}}{4\pi l^2_j}, 
\label{JLW}
\end{equation}
where $L_{\rm LW}^{k}$, the LW luminosity of the $k$-th source in the $(i,j)$-th bin, 
can take the value of one from 
$(0, L_{\rm LW}^{\rm BH}, L_{\rm LW}^{\rm gal})$, depending on $M_i$. 
When photoevaporation feedback is allowed, both DCBHs and galaxies could 
only form in halos that are massive enough to retain enough gas (we assume at least 20\%
in this paper). We use the same critical mass, $M_{\rm min}^{\rm ion}$, for both 
DCBH and galaxy formation,
as discussed in Sec. \ref{photoevaporation}. 
$L_{\rm LW}^{k}$ is therefore determined as follows:
\begin{itemize}
\item If $M_i < M_{\rm min}^{\rm ion}$, $L_{\rm LW}^{k} = 0$.  
\item If $M_i > M_{\rm min}^{\rm ion}$ and 	
$M_{\rm min}^{\rm ion}$ is 
smaller\footnote{As the probability for the gas in more massive halos to fragment 
into stars (even if metal-free) increases very rapidly
(\citealt{2009MNRAS.396..343R}, but see \citealt{2009ApJ...702L...5B}
for a different opinion), DCBH become unlikely to form above a certain mass scale which we 
take here to be $M_{\rm 2T4}$. Using a halo mass corresponding to  
$T_{\rm vir} = 5\times10^4$~K as an upper limit gives similar results. In this case even when most of gas in halos below $M_{\rm 2T4}$ has already been evaporated by ionizing photons, a few DCBHs can still form in more massive halos. However, the actual difference 
is small compared with the uncertainties of photoevaporation models.}
 than $M_{\rm 2T4}$,
for $M_{\rm min}^{\rm ion} \le M_i \le M_{\rm 2T4}$, we
suppose that a fraction $f_{\rm BH}$ of the halos in this mass range
contains active DCBHs, and for each $k$ we decide whether $L_{\rm LW}^k = L_{\rm LW}^{\rm BH}$, or 
$L_{\rm LW}^{\rm gal}$, or zero by the value of a randomly generated number, ${\cal R}_1$, following an uniform distribution:
if ${\cal R}_1 < f_{\rm BH}$, then $L_{\rm LW}^k = L_{\rm LW}^{\rm BH}$; 
if instead ${\cal R}_1 > f_{\rm BH}$, we then generate a new and independent uniformly distributed
random number ${\cal R}_2$. If ${\cal R}_2 < \Delta f_{\rm coll}/f_{\rm coll}$,
$L_{\rm LW}^k = L_{\rm LW}^{\rm gal}(M_i)$, otherwise 
$L_{\rm LW}^k = 0$. If $M_{\rm 2T4} < M_i$, the hosts only hold normal galaxies, again we decide whether 
$L_{\rm LW}^k$ equals $L_{\rm LW}^{\rm gal}$ or zero checking if a third random number ${\cal R}_3$
is smaller or larger than $\Delta f_{\rm coll}/f_{\rm coll}$.
\item If $M_i > M_{\rm min}^{\rm ion}$ and $M_{\rm min}^{\rm ion}$ is larger than 
$M_{\rm 2T4}$, we generate a random number ${\cal R}_4$; if it is smaller than 
$\Delta f_{\rm coll }/f_{\rm coll}$, $L_{\rm LW}^k = L_{\rm LW}^{\rm gal}(M_i)$,
otherwise $L_{\rm LW}^{k} = 0$.
\end{itemize}
By generating a large set of Monte Carlo realizations, we then directly estimate 
the probability that the central halo sees a LW intensity above a critical value $J_{\rm LW}^{\rm crit}$,
$p_{\rm J}$.

\end{document}